\documentclass[twocolumn,prl,showpacs,amsmath,amssymb]{revtex4}
\usepackage{graphicx}
\usepackage{epsfig}
\usepackage{dcolumn}
\usepackage{bm}

\begin{document}
\title{Quest for high $T_c$ in layered structures: the case of LiB}
\author{Matteo Calandra}
\affiliation{Institut de Min\'eralogie et de Physique des Milieux condens\'es, 
case 115, 4 place Jussieu, 75252, Paris cedex 05, France}
\author{ Aleksey N. Kolmogorov and Stefano Curtarolo}
\affiliation{Department of Mechanical Engineering and Material Science,
Duke University, Durham, North Carolina 27708, USA}
\date{\today}

\begin{abstract}
Using electronic structure calculation we study the superconducting
properties of the theoretically-devised superconductor MS1-LiB
(LiB). We calculate the electron-phonon coupling ($\lambda=0.62$) and
the phonon frequency logarithmic average ($\langle \omega
\rangle_{log}=54.6$ meV ) and show that the LiB critical temperature
is in the range of 10-15 K, despite the frozen-phonon deformation
potential being of the same order of MgB$_2$. As a consequence, LiB
captures some of the essential physics of MgB$_2$ but (i) the
electron-phonon coupling due to $\sigma$ states is smaller and (ii)
the precious contribution of the $\pi$ carriers to the critical
temperature is lacking. We investigate the possible change in $T_c$
that can be induced by doping and pressure and find that these
conditions cannot easily increase $T_c$ in LiB.
\end{abstract}
\pacs{63.20.Kr, 63.20.Dj , 78.30.Er, 74.70.Ad}

\maketitle
\section{introduction}

The quest for superconductivity in layered structures has become the
focus of intense research since the discovery of superconductivity in
MgB$_2$ (T$_c=39$ K) \cite{Nagamatsu}. The layered structure of
MgB$_2$ generates one of its most prominent features, namely the
B 2$p_{xy}$ orbitals form $\sigma$-bands \cite{An,Belashchenko,Kortus}
which are weakly dispersing along the $k_z$ direction and have a
marked two dimensional character. In MgB$_2$ the $\sigma-$bands are
hole doped, but the top of these bands is only $\approx 0.5$ eV higher
than the Fermi level. The $\sigma-$bands Fermi surface sheets
\cite{Kortus}, two slightly warped cylinders with axis perpendicular
to the boron layers, generate a huge electron-phonon coupling along
the $k_z$ direction. The carriers in the $\pi-$bands, formed by the B
2$p_z$ orbitals, further enhance the average electron-phonon coupling
\cite{Kong}.

The formation of $\sigma$ and $\pi$ states is typical of graphite-like
structures composed by boron or carbon atoms. Given the success of
MgB$_2$ it is natural to look for high $T_c$ superconductivity in
structures having similar features. The problem is that, given the
boron layers, small variations in the valence or mass of the
intercalant or in the structural parameters are sufficient to
considerably alter the $\sigma$ or $\pi$ bands positions or the shape of
their Fermi surfaces and consequently destroy superconductivity. For
one or some of these reasons AlB$_2$, ZrB$_2$, NbB$_2$, MoB$_2$,
YB$_2$, TaB$_2$, TiB$_2$, HfB$_2$, VB$_2$ and CrB$_2$ are not
superconducting \cite{Cooper,Gasparov,Leyarovska}.

The hope of finding new superconducting materials in layered
structures was recently increased by the discovery of
superconductivity in the graphite intercalated compounds, YbC$_6$ and
CaC$_6$ \cite{Weller,Genevieve}. This is particularly promising since
a huge number of intercalants are available for
graphite \cite{DresselhausRev}. In CaC$_6$, despite the layered
structure and the existence of $\sigma$ and $\pi$ bands
originated from the carbon 2$p$ orbitals, the electronic structure close
to the Fermi level is completely different from that of MgB$_2$.  
The $\pi$ bands, reminiscent of the graphite ones, and 
an intercalant free-electron-like band \cite{Csanyi,Calandra2005} 
cross the Fermi energy. 
The intercalant band forms a spherical Fermi surface
 \cite{MazinGIC,CalandraPRBGIC2006}. The electron-phonon coupling of
CaC$_6$ ($\lambda=0.83$) is mainly due to coupling of the interlayer
band with C vibrations perpendicular to the graphite layers and with Ca
vibrations. So, even though missing the $\sigma$-bands, CaC$_6$ reaches
an interesting 11.5 K $T_c$. This temperature is substantially
enhanced by pressure (T$_c=15.1$ K at $\approx 8$ GPa \cite{Gauzzi}),
contrary to what happens in MgB$_2$.

As can be seen from the above examples, even if one restricts to
sandwich structures formed by boron or carbon layers, the details of
the electronic and phonon spectra and, subsequently, the critical
temperature can change dramatically when the intercalant is
included. As a consequence, a theoretical approach is absolutely
necessary to identify the most probable superconductors or at least to
exclude the less probable ones.

An attempt in this direction has been recently made in
Ref. \cite{Rosner2002}, where by using {\it ab initio} methods the authors
studied the possible hole-doping of LiBC, a $\approx 1$ eV gap
semiconductor. The authors suggested that a $T_c$ of the order of
MgB$_2$ could be reached if the insulating LiBC is substantially doped
with holes to obtain Li$_{0.5}$BC. Successive experimental studies
have indicated that the structural response to the introduction of
holes unfavorably modifies the electronic structure of Li$_x$BC, and
so far no high $T_c$ superconductivity has been found in this
system \cite{LixBC}.

Ideally designing new superconductors {\it ab initio} requires three
steps.  The first is the determination of the most stable structures
given a set of atomic species.  The second is the calculation of the
electronic structure to verify that the given structure is at least
metallic or can be made metallic easily.  The third is the
determination of the phonon dispersion and of the electron-phonon
parameters.

The first point is a daunting task even if one restricts one's search
to a specific set of likely candidates\cite{Materials_Science}. A
systematic approach to tackle this problem has been recently offered
in the way of data mining of {\it ab initio} calculations
\cite{DataMining,Morgan2005,Curtarolo2005}. 
In this method one uses the informations obtained
from {\it ab initio} calculations of many different structures to
build a database that can be then used to judge the stability of new
structures. Application of this method to intermetallics has led to
the identification of new layered lithium monoboride phases which have
a good chance to form under proper synthesis conditions
\cite{Kolmogorov1}.

Once a stable metallic structure is given, a calculation of the phonon
spectra and of the electron-phonon coupling needs to be performed to
obtain $T_c$.  Indeed, while some qualitative information can be
extracted from electronic-structure \cite{An}, 
for a quantitative analysis step three is absolutely
necessary.

In this work we investigate the superconducting properties of the
previously determined metal sandwich (MS) lithium monoboride
\cite{Kolmogorov1} by calculating its phonon spectrum and
electron-phonon parameters. This system is metallic and, from
qualitative arguments, one can infer that $T_c$ is of the same order
of that of MgB$_2$ \cite{Kolmogorov1}. Indeed this system has an
electronic structure which is a hybrid between those of MgB$_2$ and
CaC$_6$, since there are hole-doped $\sigma-$bands forming cylindrical
Fermi surfaces and there is an intercalant band crossing the Fermi
level.  Moreover the deformation potential is comparable to that of
MgB$_2$ \cite{Kolmogorov1}. From this point of view, LiB is a much
more promising material than Li$_{0.5}$BC, because even without doping
it has a significant density of $\sigma$-states at the Fermi level.

Unfortunately, the full electron-phonon coupling calculations
performed in this paper indicate that LiB should have a $T_c$ in a
10-15 K range. We show that LiB captures some of the important physics
of MgB$_2$, namely the role of the $\sigma-$bands, but it lacks the
contribution of the $\pi$ states to the electron-phonon coupling and
it is only a far relative of CaC$_6$ because the interlayer band is
very weakly coupled with the phonons.  In an attempt to improve the
situation we examine what role the hydrostatic pressure and doping can
play in determining the critical temperature.

\section{Technical Details}

In all our calculations of the layered lithium monoboride we have used
the MS1 theoretical crystal structure basing our choice on the
following considerations. On the one hand, it has the smallest unit
cell of all MS structures, which offers computational efficiency.  On
the other hand, even though other stacking sequences are possible
(e.g. MS2, Refs. \cite{Kolmogorov1,Kolmogorov2}), MS1 is a good
representative model of the layered lithium monoboride because the
long-period shifts are expected to have little effect on its
superconducting properties \cite{MS}.  MS1-LiB has a rhombohedral unit
cell with R\={3}m space group.  There are four atoms in the primitive
unit cell with Wyckoff positions Li$(2c) (1/2-z_{\rm Li},1/2-z_{\rm
Li},1/2-z_{\rm Li})$, Li$(2c)(1/6+z_{\rm Li},1/6+z_{\rm Li},1/6+z_{\rm
Li})$, B$(2c) (-\delta,-\delta,-\delta)$ and
B$(2c)(2/3-\delta,2/3-\delta,2/3-\delta)$. The fully relaxed
parameters are $a=b=c=5.92$ \AA,
$\alpha=\beta=\gamma=29.8^{o}$.

Density functional theory (DFT) calculations are performed using the
Quantum Espresso code \cite{PWSCF} within the generalized gradient
approximation (GGA) \cite{PBE}. We use norm-conserving
pseudopotentials \cite{Troullier} with configuration $2s^{1}2p^0$ and
non-linear core correction \cite{nlcc} for Li, and configuration $2s^2
2p^1$ for B. The wavefunctions are expanded using a 50 Ry cutoff.  The
dynamical matrices and the electron-phonon coupling are calculated
using density functional perturbation theory in the linear response
\cite{PWSCF}. For the electronic integration in the phonon calculation
we use an $N_{k}=12\times12\times12$ uniform ${\bf k}$-point mesh and
Hermite-Gaussian smearing from 0.05 Ry.  For the
evaluation of the electron-phonon coupling we use an $N_k=40\times
40\times 40$ Monkhorst-Pack mesh. For the $\lambda$ average over the
phonon momentum {\bf q} we use an $N_q=4\times 4\times 4\,$ ${\bf
q}-$point mesh.  The phonon dispersion is obtained by Fourier
interpolation of the dynamical matrices computed on the $N_q$ mesh.

The pressure- and doping-induced changes in the electronic properties
of LiB are studied with Vienna Ab initio Simulation Package {\small
VASP} \cite{kresse1993,kresse1996b} within the GGA \cite{PBE}. We use
projector augmented waves (PAW)~\cite{bloechl994} pseudopotentials, in
which Li semi-core states are treated as valence; the energy cutoff is
set at 30 Ryd. The projected electronic density of states (EDOS) is
found by decomposition of the wavefunction within a sphere of the
default PAW radius of 1.7 a.u. For the MS1 unit cell, the $2\times
2\times 3$-MS1 and $2\times 2\times 1$-MS2 supercells we use $31\times
31\times 31$, $18\times 18\times 6$, and $18\times 18\times 10$
Monkhorst-Pack {\bf k}-meshes, respectively.

\section{Band-Structure, DOS and Fermi surface}

\begin{figure}[t]
\centerline{\includegraphics[width=90mm]{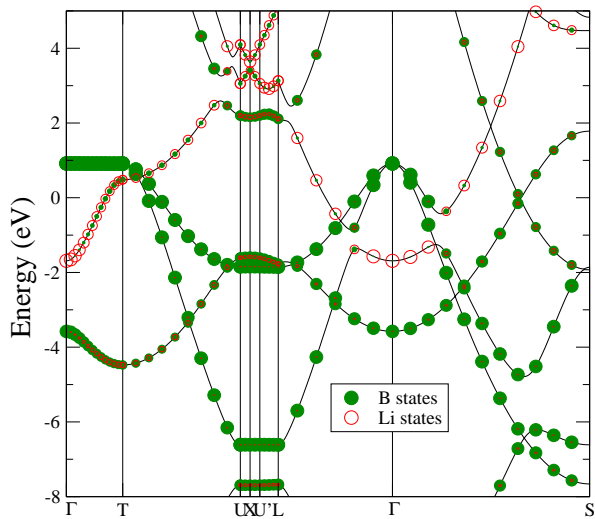}}
\caption{(Color online) Band structure of LiB. The size of the empty (full) dots
represent the amount of Li (B) character at a given {\bf k}-point.
See footnote \cite{footnote_special_k} for high-symmetry points notation.}
\label{fig:Bands}
\end{figure}

\begin{figure}[t]
\includegraphics[height=6.5cm]{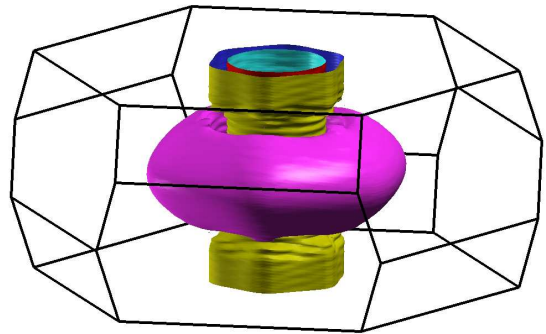}%
\caption{(Color online) Fermi surface of MS1-LiB. For convenience, the
Brillouin zone is stretched  along the
$z$-direction.}
\label{fig:FS}
\end{figure}

The band structure of Li$_{2}$B$_{2}$ is presented in
Fig. \ref{fig:Bands} (see footnote \cite{footnote_special_k} for
high-symmetry points notation). 
Similarly to what happens in MgB$_2$ \cite{An},
there are two boron $\sigma$-bands crossing the Fermi energy
$\epsilon_F$. Compared to the $\sigma$-bands in MgB$_2$, these bands
are even more two-dimensional (due to the larger interlayer distance)
and shifted by more than 0.6 eV to higher energies at the $\Gamma$
point. As in MgB$_2$, they generate two cylindrical Fermi surfaces (in
our case with axes along the $\Gamma T$ direction, Fig. \ref{fig:FS}).
The boron $\pi$ states in LiB resemble more the $\pi$ states of
graphite, as they cross exactly at $\epsilon_F$, so that LiB is
lacking $\pi$ Fermi surfaces altogether. In MgB$_2$ these states cross
at about 2 eV above $\epsilon_F$, which leads to the appearance of an
extended $\pi$ Fermi surface \cite{Kortus}. Another important
difference between the electronic structures of the two borides is the
presence of a lithium band at $\epsilon_F$ in LiB. The position of
this band resembles the intercalant band in CaC$_6$
\cite{Calandra2005}, although in LiB it has substantial hybridization
to boron states close to the $T$-point. The corresponding 
Fermi surface (a compressed sphere) is depicted in Fig. \ref{fig:FS}.

\begin{figure}[t]
\includegraphics[height=7.5cm]{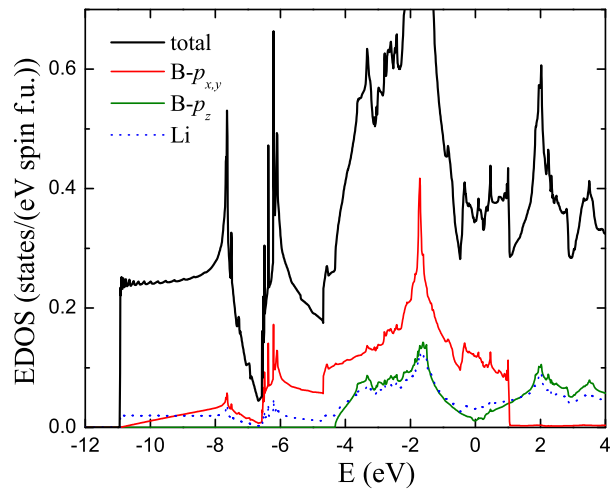}%
\caption{(Color online) Electronic density of states projected over selected atomic orbitals.}
\label{fig:DOS}
\end{figure}

The total density of states (EDOS) and the EDOS projected over atomic
orbitals is illustrated in Fig \ref{fig:DOS}. The main component at
$\epsilon_F$ is given by boron $p\sigma$ states. As in graphite the boron $p\pi$ EDOS
at $\epsilon_F$ is zero and increases slowly and linearly immediately
after $\epsilon_F$.

\section{Phonon spectrum and Superconducting Properties}

\begin{figure}[t]
\includegraphics[width=90mm]{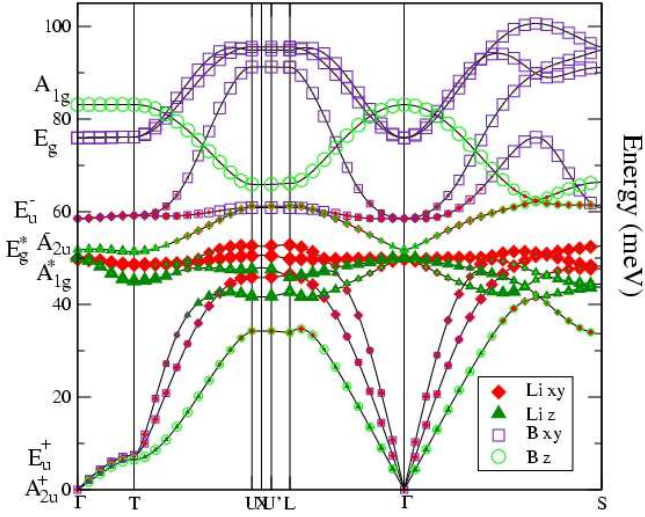}%
\caption{(Color online) Phonon dispersion of MS1-LiB with
decomposition in in plane Li and B vibrations (labeled Li$_{xy}$ and
B$_{xy}$ respectively) and out-of-plane Li and B vibrations (labeled
Li$_{z}$ and B$_{z}$ respectively).  Notation for the phonon modes at
the $\Gamma$-point, given next to the vertical axis, is explained in
the text.}
\label{fig:phonon_dispersion}
\end{figure}

The phonon dispersion and the phonon density of states (PHDOS)
are illustrated in Fig.
\ref{fig:phonon_dispersion}. The phonon modes at the $\Gamma$-point
are decomposed as $2A_{1g} + 2A_{2u} + 2E_g + 2E_u$\cite{MODES}. To
distinguish between modes with the same symmetry but different
eigenvectors we use the following notation:
$A_{1g}    $ $[{\rm B}_z]$,
$E_g       $ $[{\rm B}_{xy}]$,
$E_u^{-}   $ $[({\rm Li}-{\rm B})_{xy}]$,
$A_{1g}^*  $ $[{\rm Li}_z]$,
$A_{2u}^{-}$ $[({\rm Li}-{\rm B})_{z}]$,
$E_g^*     $ $[{\rm Li}_{xy}]$,
$A_{2u}^{+}$ $[({\rm Li}+{\rm B})_{z}]$,
$E_u^{+}   $ $[({\rm Li}+{\rm B})_{xy}]$,
where in brackets we give the corresponding atoms and vibrations.
For convenience, we label phonon branches everywhere in the Brillouin
zone using the name of their representation at $\Gamma$.

Except for acoustic modes, a clear separation exists between optical
Li and B vibrations. Li modes are confined in the 40-55 meV region and
are not dispersive, meaning that Li-vibrations behave essentially as
Einstein modes. Boron in-plane vibrations are softened along the
$\Gamma T$ direction due to coupling to the $\sigma$ bands. The
softening at $\Gamma$ of the $E_g$ phonon branches is approximately 20
meV, to be compared with the almost 40 meV in MgB$_2$ for the $E_{2g}$
modes \cite{Kong,Shukla}. This suggests a strong coupling of the
$\sigma$ bands to the in-plane vibrations in LiB \cite{Kolmogorov1},
but not as strong as in the case of MgB$_2$.

The three acoustic modes along the $\Gamma T$ direction are
substantially softer with respect to the other directions. At the zone
border, $T$, these modes are formed by (i) in-plane (Li+B)$_{xy}$
vibrations at energies $\approx 8.5$ meV and (ii) out-of-plane
(Li+B)$_z$ vibrations at $\approx 8.0$ meV. They can be related to
the soft modes at the $\Gamma$ point in MS2-LiB, discussed in Ref. \cite{Kolmogorov1}.

The superconducting properties can be understood calculating the
electron-phonon coupling $\lambda_{{\bf q}\nu}$ for a phonon mode
$\nu$ with momentum ${\bf q}$:
\begin{equation}\label{eq:elph}
\lambda_{{\bf q}\nu} = \frac{4}{\omega_{{\bf q}\nu}N(0) N_{k}} \sum_{{\bf k},n,m} 
|g_{{\bf k}n,{\bf k+q}m}^{\nu}|^2 \delta(\epsilon_{{\bf k}n}) \delta(\epsilon_{{\bf k+q}m})
\end{equation}
where the sum is over the Brillouin Zone.
The matrix element is
$g_{{\bf k}n,{\bf k+q}m}^{\nu}= \langle {\bf k}n|\delta V/\delta u_{{\bf q}\nu} |{\bf k+q} m\rangle /\sqrt{2 \omega_{{\bf q}\nu}}$,
where $u_{{\bf q}\nu}$ is the amplitude of the displacement of the phonon, 
 $V$ is the Kohn-Sham potential and $N(0)$ is the electronic density of states 
at the Fermi level.
The calculated average electron-phonon coupling is  
$\lambda=\sum_{{\bf q}\nu} \lambda_{{\bf q}\nu}/N_q\approx0.62$
($N_{k}$ and $N_{q}$ are the previously defined ${\bf k}$-space and ${\bf q}$-space mesh dimensions, respectively).

\begin{figure}[t]
\includegraphics[width=10.0cm]{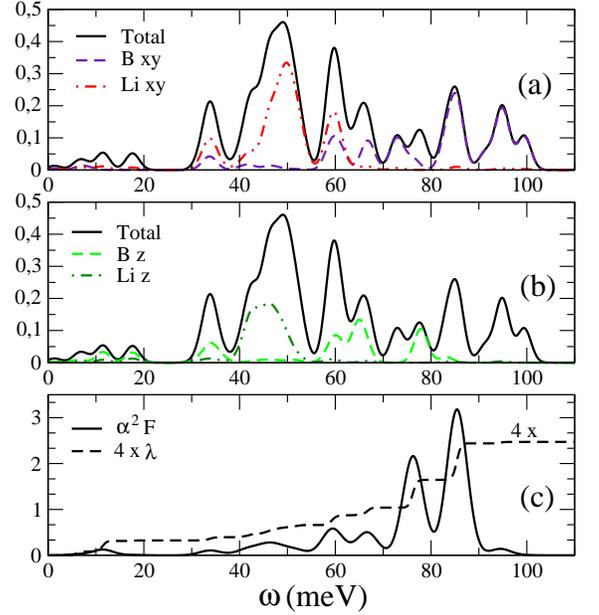}%
\caption{(Color online) Phonon density of states (PHDOS),
partial phonon density of states projected over selected
vibrations, Eliashberg function and integrated Eliashberg function for MS1-LiB.
For clarity, the integrated Eliashberg function has been multiplied by a factor
of 4}
\label{fig:alpha2fDOS}
\end{figure}

The Eliashberg function
\begin{equation}
\alpha^2F(\omega)=\frac{1}{2 N_q}\sum_{{\bf q}\nu} \lambda_{{\bf q}\nu} \omega_{{\bf q}\nu} \delta(\omega-\omega_{{\bf q}\nu} )
\end{equation}
and the integral $\lambda(\omega)=2 \int_{0}^{\omega} d\omega^{\prime} 
\alpha^2F(\omega^{\prime})/\omega^{\prime}$ 
are shown in Fig. \ref{fig:alpha2fDOS}.
As can be seen most of the contribution comes from phonon states in the 60-90 meV
region and a smaller contribution comes from low energy states.

An estimate of the different contributions of the in-plane (Li$_{xy}$ and B$_{xy}$)
and out-of-plane (Li$_z$ and B$_z$) vibrations 
to $\lambda$ can be obtained from the relation
\begin{equation}\label{eq:trlambda}
\lambda=\sum_{i\alpha j\beta}\Lambda_{i\alpha, j\beta}=
\sum_{i\alpha j\beta}\frac{1}{N_q}\sum_{\bf q}
 [{\bf G}_{\bf q}]_{i\alpha,j\beta} [{\bf C_q}^{-1}]_{j\beta,i\alpha}
\end{equation}
where $i,\alpha$ indices indicate the displacement of the $i^{\rm th}$ atom in the Cartesian direction $\alpha$, 
$[{\bf G_q}]_{i\alpha,j\beta}=\sum_{{\bf k},n,m}4 {\tilde
g}_{i\alpha}^{*}{\tilde g}_{j\beta} \delta(\epsilon_{{\bf k}n})
\delta(\epsilon_{{\bf k+q}m})/[N(0) N_{k}]$, and ${\tilde
g}_{i\alpha}=\langle {\bf k}n|\delta V/\delta x_{{\bf q} i\alpha}
|{\bf k+q} m\rangle /\sqrt{2}$.  The ${\bf C_q}$ matrix is the Fourier
transform of the force constant matrix (the derivative of the forces
with respect to the atomic displacements).  The decomposition leads
to:
\begin{equation}
  \bm{\Lambda}\,=
  \begin{matrix}
    &
    \begin{matrix}
      {\rm B}_{xy} &  {\rm B}_{z} & {\rm Li}_{xy} & {\rm Li}_z \\
    \end{matrix} \\
    \begin{matrix}
       {\rm B}_{xy} \\
       {\rm B}_{z}  \\
       {\rm Li}_{xy}\\
       {\rm Li}_z   \\
    \end{matrix} &
    \begin{pmatrix}
   .46  &   .00 &   -.02  &   .00   \\
   .00  &   .13 &    .02  &  -.05   \\
  -.02  &   .02 &    .08  &  -.01   \\
   .00  &  -.05 &   -.01  &   .07   \\ 
    \end{pmatrix}
  \end{matrix}
\end{equation}

The off-diagonal terms are small (but not negligible) compared to the
total $\lambda$. Most of the coupling is to the in-plane B vibration;
contributions from the Li and the out-of-plane B vibrations are
smaller. Since the $\sigma$-bands do not couple to the B$_z$
vibrations and since there are no $\pi$ Fermi surfaces, the coupling
to B$_z$ vibrations is due to the intercalant band. Note that the
decomposed values of $\lambda$ contain contributions from different
modes and are summed over all the ${\bf q}$-points in the Brillouin
zone. For example, $\Lambda_{{\rm B}_{xy},{\rm B}_{xy}}=0.46$ includes
the coupling to the in-plane $E_g$, $E_u^-$, and $E_u^+$ branches. By
examining the integrated Eliashberg function $\lambda(\omega)$ in
Fig. \ref{fig:alpha2fDOS}(c) and the phonon characters in
Fig. \ref{fig:phonon_dispersion} one can infer that the $E_g$ branch
is the most important of the three: among them it has the highest
PHDOS in the 70-100 meV range, in which $\lambda$ gains most of its
total value. The soft in-plane $E_u^+$ branch is far less important,
as the net contribution from all the soft modes having energy under 20
meV is only $\approx 0.08$ (Fig. \ref{fig:alpha2fDOS}(c)).

It is instructive to compare our result with other layered
superconductors. In MgB$_2$ the coupling of the $\sigma$-bands to the
phonon modes is $\lambda^{MgB_2}_{\sigma,\sigma}=0.62\pm 0.05$
\cite{Kong}, while in LiB the corresponding value is less than 0.46,
as discussed above.  This difference can be clarified by noting that
the $E_{2g}$ phonon linewidth $\gamma_{{\bf q},E_{2g}}=2\pi N(0)
\omega_{{\bf q}E_{2g}}^2\lambda_{{\bf q}\nu}$ along $\Gamma A$ in
MgB$_2$ happens to be comparable in magnitude with that of the $E_g$
mode along $\Gamma T$ in LiB. Therefore, the reduced electron-phonon
coupling in LiB is mainly due to the $E_g$ phonon frequency being
harder than the $E_{2g}$ one in MgB$_2$
($\omega^{LiB}_{E_g}/\omega^{MgB_2}_{E_{2g}}\approx 1.3$ at $\Gamma$).
This unfortunate result can be linked to the absence of the $\pi$
carriers, which play an important role in softening of the $E_{2g}$
mode in MgB$_2$\cite{Mazin_Antropov,Liu2006}.

We find that LiB and graphite intercalated compounds have few
similarities in terms of superconducting features. In particular, in
CaC$_6$ the intercalant modes are responsible of $\sim 50\%$ of the
total electron-phonon coupling, and the rest comes from vibrations of
carbon modes in the direction perpendicular to the graphite layers.
In CaC$_6$ one has $\lambda_{\rm Ca_{xy}} + \lambda_{\rm Ca_{z}} =
0.33$ and $\lambda_{\rm C_{z}}=0.33$ \cite{Calandra2005} .
In LiB the overall contribution of B$_z$, Li$_{xy}$ and
Li$_{z}$ vibrations is less than half of that of CaC$_6$, which means
that while LiB captures some of the physics of MgB$_2$, it does not
capture the physics of graphite intercalated compounds to full
extent.
This is also clear from the phonon spectrum of CaC$_6$ where the
intercalant modes are at energies lower than 20 meV and one of the Ca
modes undergoes a marked softening with a corresponding large
electron-phonon coupling (at point $X$ of Fig. 2 in
Ref. \cite{Calandra2005}). In LiB, on the contrary, the Li modes are
much higher in energies ($\sim 50$ meV) meV and dispersionless. The
main reason for this difference comes from the mass of Li which is
5.77 times smaller than that of Ca leading to frequencies which are on
average 2.4 times larger.

The critical superconducting temperature is estimated using the McMillan 
formula \cite{mcmillan}:
\begin{equation}
T_c = \frac{\langle \omega \rangle_{log}}{1.2}\, \exp \left[{ - \frac{1.04 (1+\lambda)}{\lambda-\mu^* (1+0.62\lambda)}}\right]
\label{eq:mcmillan}
\end{equation}
where $\mu^*$ is the screened Coulomb pseudopotential and
\begin{equation} 
\langle\omega\rangle_{log} = e^{\frac{2}{\lambda}\int_{0}^{+\infty} 
\alpha^2F(\omega)\log(\omega)/\omega\,d\omega }
\end{equation}
the phonon frequencies logarithmic average. We obtain $\langle\omega\rangle_{log}=54$ meV
leading to $T_c$ of approximately 10-15 K for $\mu^{*}=0.14-0.1$. This value could
be further enhanced by multiband effects.

\section{Doping and pressure effects}
\label{sec:Pdelta}

Even though a theoretically-devised from scratch superconductor with
$T_c=10-15$ K could be considered a success of the materials
prediction methodology, the stoichiometric LiB compound falls short of
expectations to compete with the record-holding binary MgB$_2$. In
this section we investigate whether it is possible to favorably modify
the electronic properties of LiB and achieve higher $T_c$ by doping or
applying pressure. We pay special attention to the evolution of the
$\pi$ states, since their reintroduction at $\epsilon_F$ may soften
the $E_g$ mode and lead to a larger coupling.

\begin{figure}[t]
\centerline{\includegraphics[width=9.5cm]{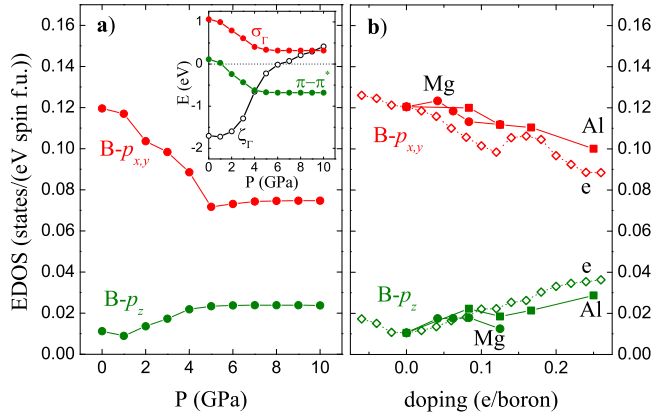}}%
\caption{(Color online) a) EDOS at the Fermi level projected over
$\sigma$ and $\pi$ states as a function of pressure; b) the same as a
function of doping level: the hollow points represent charged LiB unit
cell in a neutralizing background; the solid points correspond to
substitutional Mg and Al doping of LiB in the MS1 and MS2
superstructures described in the text \cite{details}. For comparison,
the $\sigma$ and $\pi$ EDOS in MgB$_2$ are 0.098 and 0.064
states/(eV$\cdot$spin$\cdot$f.u.), respectively. The inset shows
pressure-induced changes in the position of different states near the
Fermi level: $\sigma$-boron and intercalant (labeled $\zeta$) states
at the $\Gamma$ point and crossing of the $\pi$-$\pi^*$ states along
the $\Gamma S$ direction.}
\label{fig:edosPdelta}
\end{figure}

As has been pointed out previously \cite{Kolmogorov1}, the bonding
$\pi$ states are completely filled under ambient conditions. Because
the band crossing in LiB at the Fermi level is accidental, it may be
possible to move the crossing point with pressure and increase the
$\pi$-bands EDOS at $\epsilon_F$. Figure \ref{fig:edosPdelta}(a)
reveals that there is indeed a rapid change in the $\sigma$ and $\pi$
EDOS, followed by a plateau after 5 GPa. This behavior is a reflection
of two distinctly different regimes of the LiB structural changes: i)
in the 0-5-GPa pressure range the Li-Li interlayer spacing quickly
shrinks and the B-B bond length slightly expands so that at 5 GPa they
become about 0.5 and 1.02 of their zero-pressure values, respectively;
ii) for pressures above 5 GPa the Li atoms in the bilayer start
experiencing the hard-core repulsion and the compound compresses more
isotropically. The inset in Fig. \ref{fig:edosPdelta}(a) illustrates
that by reducing the Li-Li interlayer distance one forces the charge
from the intercalant band (completely emptied at about 6 GPa) into the
boron $\pi$ and $\sigma$ states (lowered by 0.7 eV at that
pressure). Once the charge redistribution is complete, no appreciable
changes in the EDOS are seen for the boron states up to at
least 30 GPa. Therefore, the peculiar behavior of the nearly free
electron intercalant states (also observed in other systems
\cite{Csanyi,NFE}) is the only meaningful factor allowing modification
of the LiB boron states with pressure.

These simulations demonstrate that the compression of LiB does not
lead to the desired $\pi$-bands EDOS values comparable to those in
MgB$_2$. Moreover, the hydrostatic pressure causes such a quick drop
in the $\sigma$-bands EDOS that this will likely negate any possible
enhancement in the electron-phonon coupling due to the reintroduction
of the $\pi$-states at $\epsilon_F$. The phonon modes are also
expected to harden under pressure, further reducing the
electron-phonon coupling in LiB \cite{softening}.

LiB has plenty of available bonding $\sigma$ states, therefore the
compound should be easy to electron-dope. A quick examination of the
boron EDOS states around the Fermi level (Fig. \ref{fig:DOS} or Fig. 4
in Ref. \cite{Kolmogorov1}) gives an idea on what possible changes in
the Fermi surfaces and, eventually, in the electron-phonon coupling
the doping could lead to. At small doping levels, when the rigid band
approximation normally holds, the EDOS from the two dimensional
$\sigma$ bands should only slightly fluctuate until the states are
completely filled, which happens at $\Delta q\approx e $N(0)$ \Delta
E\approx 0.70$ (e/eV) (1 eV) = 0.70 e/f.u. = 0.35 e/boron. The EDOS
from the $\pi$ bands grows slowly and even at the relatively high
electron-doping of 0.35 e/boron it would amount only to about a half
of what is observed in MgB$_2$. Another way to tweak the electronic
structure could be to hole-dope LiB as it is done for Li$_x$BC
\cite{LixBC}. The known limitations of this approach are the buckling
of the hexagonal layers and the eventual destabilization of the
compound upon heavy Li depletion \cite{LixBC}.

We first simulate the electron- (hole-) doping using a charged cell
with a neutralizing positive (negative) background. Normally, in this
approach one can safely relax the unit cell parameters and obtain
valuable information about the bond length variation under small
doping.  However, in the case of the electron-doped LiB the repulsion
between the negatively charged boron layers overcomes the weak binding
between the lithium layers, causing the $c$-axis to undergo unphysical
expansion even at small levels of doping. Therefore, we fix the
$c$-axis at the zero-doping value and relax only the remaining three
parameters. The set of data, shown as hollow points in
Fig. \ref{fig:edosPdelta}(b), supports our earlier conclusion that the
$\pi$-band EDOS cannot be easily increased. Note that the
approximations used in this test, i.e. the fixed $c$-axis and the use
of a neutralizing background, may influence the results to some
extent. For example, the positive electrostatic potential from ionized
dopants could bring the delocalized $\pi$-states down (in addition to
the rigid band downshift) and could potentially be an important factor
in increasing the $\pi$ EDOS.

To address these limitations we use a more realistic model of the
electron-doped LiB by substituting Li with Mg or Al. Small doping
levels are obtained only for large unit cells; we use the hexagonal
$2\times 2\times 3$-MS1 and $2\times 2\times 1$-MS2 supercells with 48
and 32 atoms, respectively. Replacement of one or two Li atoms in
these structures results in the 1/24, 1/16, 1/12, and 1/8
concentrations of dopants per boron, and the level of doping is found
in the assumption that they give up all their valence charge. In all
the cases the $c$-axis expansion in the fully relaxed unit cells does
not exceed 6\%. The resulting averaged boron EDOS for the Mg and Al
sets are shown in Fig. \ref{fig:edosPdelta}(b) as solid points
\cite{details}. The scattered presence of dopants in the lattice
should cause some dispersion of the local boron properties. A general
trend observed in our supercell calculations is that a downshift of
the $\pi$ and $\sigma$ states happens only for B layers in direct
contact with the dopant. Typical values of the downshift that a single
Mg (Al) atom induces in all eight atoms in a neighboring B layer are
about 0.2 (0.5) eV. It is not easy to isolate the importance of
different factors defining the level of B doping, i.e. the simple
charge transfer, the electrostatic effect discussed above and the
structural changes (expansion of the $c$-axis and contraction of the
B-B bond). However, Fig. \ref{fig:edosPdelta}(b) demonstrates that the
net effect of the substitutional doping is described reasonably well
within the rigid band model.

In summary, our tests indicate that it is rather difficult to
reintroduce a significant amount of $\pi$ states at $\epsilon_F$ with
hydrostatic pressure or small doping, because the band crossing in LiB
happens to be exactly at $\epsilon_F$, about 2 eV lower than in
MgB$_2$. To have a chance of substantially increasing $T_c$, one
should search for more radical ways of modifying the electronic
structure of the MS metal borides.

\section{Conclusions and perspectives}

In this work we have investigated electron and phonon properties of
the recently theoretically-devised superconductor LiB
\cite{Kolmogorov1}. By studying in details the phonon properties of
this hypothetical material we have found that its critical temperature
is of the order of 10-15 K. Superconductivity
is mainly of the MgB$_2$ kind with planar $p\sigma$ bands strongly
coupled with phonons.  Differently from MgB$_2$, LiB has no Fermi
surface generated by $\pi$ states but an additional intercalant one.
Thus, its electronic structure can be seen as a hybrid between MgB$_2$
and CaC$_6$. However, the intercalant electronic states of LiB are
weakly coupled with the Li and B$_z$ vibrations so that the overall
electron-phonon coupling is only $\lambda=0.62$.

Since the discovery of MgB$_2$, no other diborides have been found
with high $T_c$ (for a full list, see Ref. \cite{Buzea}). If we
compare LiB with the known diborides, our calculated 10-15 K critical
temperature is not so low, although it is far from the 39 K of
MgB$_2$. Nevertheless, the study of LiB gives an important
understanding. The common belief is that the main effect for the
singular and unique behavior of MgB$_2$ is given by the presence of
almost two-dimensional $\sigma$ bands.  LiB has even more planar
$\sigma$ bands and even higher $\sigma$ EDOS at $\epsilon_F$ relative
to that in MgB$_2$ \cite{Kolmogorov1,Kolmogorov2}; however, their
contribution to the total electron-phonon coupling turns out to be at
least 25\% smaller. This reduction can be attributed to the
differences between the in-plane boron vibrations in the two borides,
caused mainly by the lack of the $\pi$ carriers at $\epsilon_F$ in
LiB. Namely, the softening of the $E_g$ mode in LiB is substantially
smaller (by about a factor of two) than that of $E_{2g}$ in MgB$_2$,
which makes the former mode be noticeably harder
($\omega^{LiB}_{E_g}/\omega^{MgB_2}_{E_{2g}}\approx 1.3$ at
$\Gamma$). Therefore, in addition to the direct loss of the $\pi$
states contribution to $\lambda$, their absence at $\epsilon_F$ in LiB
also has a strong indirect negative effect on the overall
electron-phonon coupling.

We have investigated whether this somewhat unexpected obstacle,
preventing LiB to be a truly high $T_c$ superconductor, could be
overcome with moderate modifications of the compound's
properties. Behavior of electronic features important for the LiB
superconductivity has been examined as a function of small doping and
pressure. Our results indicate that the $\pi$ EDOS increases very
slowly and cannot reach the desired values, at least not before the
$\sigma$ EDOS is substantially reduced. Thus, small doping and
pressure are not expected to significantly improve $T_c$ in LiB. A
promising direction to fix the problem would be to find a suitable
LiB-based ternary alloy; this question is currently under
investigation.

Theoretical development of potentially important superconducting
materials is a difficult task because $T_c$ critically depends on
their band structure features and vibrational properties. The
challenge is even greater if one attempts to design a superconductor
from scratch, since one first needs to ensure its thermodynamic
stability. The case of LiB shows that it is possible to theoretically
predict a compound that both i) has a good chance to form and ii)
possesses interesting superconducting properties. Study of such
promising candidates gives important insights into how to perform a
more targeted search for novel superconducting materials.

While we were finishing writing this paper, a preprint on the related
structure MS2-LiB appeared on-line \cite{Liu2006}.  The results of the
paper are similar to ours except for some numerical details that can
probably be related to the different unit cells and {\bf k},{\bf
q}-point samplings used\cite{footnote}.  However, our conclusions concerning the
possibility of increasing the $T_c$ in LiB by doping are rather
different, as explained in the previous Section.

\section{Acknowledgments}

We acknowledge many fruitful discussions with Igor Mazin, 
Francesco Mauri, Michele Lazzeri, and Roxana Margine. 
Calculations were performed at the San
Diego supercomputing center and at IDRIS supercomputing center
(project 061202).


\begin{thebibliography}{99}
\bibitem{MS} For example, the long-period shifts have little effect on
the MS-LiB energy: MS1 and MS2 have been demonstrated to be nearly
degenerate under ambient conditions, so that the synthesized lithium
monoboride could be a mixture of the two crystal structure
phases \cite{Kolmogorov1}.
\bibitem{footnote_special_k}
The special points are: $\Gamma=(0,0,0)$,T$=(1/2,1/2,1/2)$,
U$=(1/2+\eta,\tau,\tau)$,X$=(1/2,0,0)$, U$^{'}=(1/2-\eta,-\tau,-\tau)$,
L$=(1/4,-1/4,-1/4)$,S$=(1/2,-1/2,0)$. Given the 
rhombohedral angle $\alpha$, the parameters $\eta$ and 
$\tau$ are expressed as $\eta=\left[1-2/(3\tan^2(\alpha/2)-1)\right]/12$
 and $\tau=\left[1+1/(3\tan^2(\alpha/2)-1)\right]/12$.
\bibitem{softening} We do not observe any significant softening of the
in-plane boron phonon mode resulting from the ~2\% expansion of the
boron-boron bond under small (below 5 GPa) pressures. The frequencies
of the soft sliding modes have been shown to nearly double at 2
GPa \cite{Kolmogorov1}.
\bibitem{details} We place the dopants in the supercell as uniformly
as possible. Due to the high sensitivity of the projected EDOS to
k-sampling we use a 0.2 eV smearing in the MS1 and MS2 supercell
calculations.
\end{thebibliography}
\end{document}